# Numerical investigation of the formation and stability of homogeneous pairs of soft particles in inertial microfluidics

Benjamin Owen[1] and Timm Krüger[1]

[1]School of Engineering, Institute for Multiscale Thermofluids, University of Edinburgh, Edinburgh, UK

**Abstract**

We investigate the formation and stability of a pair of identical soft capsules in channel flow under mild inertia. We employ a combination of the lattice Boltzmann, finite element and immersed boundary methods to simulate the elastic particles in flow. Validation tests show excellent agreement with numerical results obtained by other research groups. Our results reveal new trajectory types that have not been observed for pairs of rigid particles. While particle softness increases the likelihood of a stable pair forming, the pair stability is determined by the lateral position of the particles. A key finding is that stabilisation of the axial distance occurs after lateral migration of the particles. During the later phase of pair formation, particles undergo damped oscillations that are independent of initial conditions. These damped oscillations are driven by a strong hydrodynamic coupling of the particle dynamics, particle inertia and viscous dissipation. While the frequency and damping coefficient of the oscillations depend on particle softness, the pair formation time is largely determined by the initial particle positions: the time to form a stable pair grows exponentially with the initial axial distance. Our results demonstrate that particle softness has a strong impact on the behaviour of particle pairs. The findings could have significant ramifications for microfluidic applications where a constant and reliable axial distance between particles is required, such as flow cytometry.

## 1 Introduction

Microfluidic devices play an increasingly important role in disease diagnostics. Due to their small footprint, high portability, relatively low cost and ever-improving manufacturing techniques, microfluidic devices have the potential to revolutionise point-of-care applications. Microfluidic devices exploit physical effects, such as flow and cell dynamics, at a smaller scale than conventionally sized devices. For example, cells can be manipulated into forming regularly spaced pairs and trains that are required for applications such as flow cytometry (Hur *et al.*, 2010) and cell encapsulation (Moon *et al.*, 2018).

Inertial microfluidics (IMF) is a relatively new research field that emerged in the late 2000s (Di Carlo *et al.*, 2007; Russom *et al.*, 2009). While conventional microfluidic devices are mostly operated in the limit of small Reynolds number, the Reynolds number in IMF is typically of the order of 10–100 due to relatively high flow speeds. Besides an increased throughput, IMF features additional physical effects that can be exploited for particle manipulation, focussing in the channel cross-section, and separation. At finite inertia, particles and cells experience additional lift forces which lead to cross-streamline migration of particles that would not migrate in the Stokes limit (Ho & Leal, 1974; Schonberg & Hinch, 1989; Asmolov, 1999; Matas *et al.*, 2009). Inertial effects can also lead to axial ordering of particles (Matas *et al.*, 2004; Lee *et al.*, 2010), an effect that can be exploited to overcome the Poisson statistics that is often plaguing applications in the non-inertial regime (Lagus & Edd, 2013). The ultimate aim is to bring IMF to maturity and routinely use it for diagnostic applications, such as recovery of rare cells from blood (Tanaka *et al.*, 2012), separation of particles by deformability (for instance, diseased red blood cells from healthy ones) or search for sepsis markers (Gossett *et al.*, 2012).

When the Reynolds number is of the order of 10–100, rigid spherical particles in tube flow migrate to a radial equilibrium position about 60% away from the tube centreline. This so-called Segré-Silberberg effect (Segre &

Silberberg, 1961) is caused by a balance of inertial shear-gradient lift forces, pushing the particles towards the wall, and wall repulsion forces caused by an increased pressure between the particles and the wall (Ho & Leal, 1974; Schonberg & Hinch, 1989; Asmolov, 1999; Matas *et al.*, 2009). The resulting lateral motion of the particles towards their equilibrium position is termed inertial migration. Altering the channel cross-section (Kim *et al.*, 2016) or curving the channel into serpentine (Zhang *et al.*, 2014) or spiral (Warkiani *et al.*, 2014) geometries changes the number and location of lateral equilibrium positions and can accelerate lateral focussing of particles (Martel & Toner, 2014).

Soft particles give rise to more complex behaviour than rigid particles since particle softness alone leads to cross-streamline migration, even in the non-inertial limit (Chen, 2014). Additionally, the analysis of soft particle dynamics is more difficult due to the changing particle shape. The inertial migration of a single soft particle has been investigated extensively in a variety of different flow conditions, including shear flows (Ma *et al.*, 2019), channel flows (Coclite *et al.*, 2020), stratified flows (Jyothi *et al.*, 2019) and viscoelastic flows (Ni & Jiang, 2020). Hur *et al.* (2011) identified experimentally that softer particles migrate to equilibrium positions closer to the channel centre. Kilimnik *et al.* (2011) also demonstrated that the equilibrium position of soft particles is essentially independent of Reynolds number, a result confirmed by Schaaf & Stark (2017).

The situation gets even more interesting when particles are sufficiently close and interact hydrodynamically. In inertial microfluidics, stable pairs and trains of particles can form under some circumstances, which brings up important implications for particle manipulation, focussing and sorting. Lee *et al.* (2010) first identified the self-assembly of particle pairs in inertial flows through the mechanism of reversing streamlines. Since then, a number of studies have investigated the pair/train formation further, both numerically and experimentally. For stable pairs, axial distance has been shown to be independent of initial position for rigid particle pairs (Humphry *et al.*, 2010; Schaaf & Stark, 2020), while different trajectory types have been identified, depending on the interaction between the particles. Of these identified trajectory types, only damped oscillation trajectories have been shown to lead to stable pairs (Lan & Khismatullin, 2014; Schaaf *et al.*, 2019). Kahkeshani *et al.* (2016) observed the existence of two different axial equilibrium distances between rigid pairs, with the preference to each being Reynolds number dependent. Importantly, Schaaf & Stark (2020) identified that during the formation of particle trains, stable particle pairs first form and then group together to build particle trains. Patel & Stark (2021) investigated the effect of pair softness and shape for mono- and bi-disperse pairs, finding that increased particle softness lead to increased pair stability. However, the formation mechanisms of particle pairs and the conditions leading to stable pairs are still not well understood. Moreover, the distinct properties of soft particle pairs, compared to rigid particle pairs, need to be explored further to improve the design of inertial microfluidic devices that are used for soft cells.

In this paper, we investigate the dynamics of a pair of identical soft capsules in straight channel flow at moderate inertia via immersed-boundary-lattice-Boltzmann-finite-element simulations (section 2). We validate the model by simulating the interaction of a pair of soft capsules in a simple shear flow and the lateral migration of a single soft capsule in channel flow (section 3). Our results (section 4) show that pairs of identical soft capsules exhibit six different trajectory types, two of which have not been observed for rigid pairs. We demonstrate that the lateral equilibrium positions of the particles largely determine the stability of the pair and that the stabilisation of the axial distance between both particles occurs only after the lateral migration phase. These observations lead to the hypothesis that certain mechanisms contribute more strongly to the overall system: flow development as zeroth-order, single particle lateral migration as first-order, and axial spacing as second-order effects. We also observe two distinct phases of pair formation, an early axial approach phase and a later spiralling convergence phase that is largely independent of the initial phase. During the spiralling phase, particles are tightly coupled through hydrodynamic interactions. The spiralling dynamics is characterised by frequency and damping coefficients that are determined by particle softness, while the pair formation time is mostly dependent on the initial positions of the particles. We argue that our findings have strong implications for the understanding of particle interactions in mildly inertial flows and the design of inertial microfluidic devices aiming at the formation of regularly spaced particle pairs and trains (section 5).

## 2  Physical and Numerical Model

The physical and numerical models are briefly outlined in sections 2.1 and 2.2, respectively.

## 2.1 Physical Model

### 2.1.1 Governing equations and physical parameters

We consider a single or two soft capsules flowing in a simple planar shear flow or in a straight channel. We assume an incompressible Newtonian liquid. The suspended particles are hyperelastic and neutrally buoyant capsules which are filled with the same liquid and are spherical in their undeformed state. While the liquid is governed by the incompressible Navier-Stokes equations, we employ two different elastic models for the capsules, either the Skalak model (Skalak *et al.*, 1973)

$$w_s = \frac{\kappa_s}{12}(I_1^2 + 2I_1 - 2I_2) + \frac{\kappa_\alpha}{12}I_2^2 \tag{1}$$

or the neo-Hookean model

$$w_s = \frac{\kappa_s}{6}\left(I_1 - 1 + \frac{1}{I_2 + 1}\right) \tag{2}$$

where $w_s$ is the areal energy density, $I_1$ and $I_2$ are the in-plane strain invariants (Krüger *et al.*, 2011), and $\kappa_s$ and $\kappa_\alpha$ are the elastic shear and area dilation moduli. We include a membrane bending energy

$$w_b = \frac{\kappa_b}{2}\left(H - H^{(0)}\right)^2 \tag{3}$$

where $H$ and $H^{(0)}$ are the trace of the surface curvature tensor and the spontaneous curvature, respectively, and $\kappa_b$ is the bending modulus.

The two flow scenarios considered are (i) simple shear flow between two flat and rigid plates and (ii) force-driven flow in a straight and rigid channel with rectangular cross section. The no-slip boundary condition is assumed at the surfaces of the channel and the particles.

The relevant parameters are liquid density $\rho$ and kinematic viscosity $\nu$; the radius $a$ of the undeformed capsules; the elastic shear modulus $\kappa_s$, bending modulus $\kappa_b$ and area dilation modulus $\kappa_\alpha$ of the capsules; the channel half-width $w$, half-height $h$ and length $L$; and either shear rate $\dot\gamma$ for simple shear flow or maximum velocity $U_{\max}$ at the channel centre for force-driven flow.

### 2.1.2 Dimensionless groups

The particle Reynolds number $\mathrm{Re}_p$ is used for the simple shear flow cases and defined as by Doddi & Bagchi (2008):

$$\mathrm{Re}_p = \frac{\dot\gamma a^2}{\nu}. \tag{4}$$

For the channel flow cases, the channel Reynolds number $\mathrm{Re}_c$ is used. We adopt the definition of Schaaf & Stark (2017); Schaaf *et al.* (2019):

$$\mathrm{Re}_c = \frac{U_{\max} w}{\nu}. \tag{5}$$

The capillary number Ca is the ratio of the viscous stress of the liquid to the characteristic elastic shear stress of the capsule membrane:

$$\mathrm{Ca} = \frac{\rho\nu\dot\gamma a}{\kappa_s}. \tag{6}$$

While the capillary number depends on the flow field, the Laplace number La is a combination of material properties only and suitable to isolate the contribution of particle softness in inertial flows (Schaaf & Stark, 2017):

$$\mathrm{La} = \frac{\kappa_s a}{\rho\nu^2}. \tag{7}$$

Other dimensionless groups are the confinement ratio $\chi = a/h$ or $\chi = a/w$, depending on whether $h$ or $w$ is smaller, the channel aspect ratio $\alpha = w/h$, the reduced dilation modulus $\tilde\kappa_\alpha = \kappa_\alpha/\kappa_s$ and the reduced bending modulus $\tilde\kappa_b = \kappa_b/(\kappa_s a^2)$.

### 2.1.3 Characteristic scales

In order to non-dimensionalise reported distances (or positions) and times, we use characteristic length scales and a time scale. Depending on context, distances or locations are non-dimensionalised either by particle radius $a$, channel half-width $w$, or channel half-height $h$. For channel flow, time is non-dimensionalised by the advection time

$$t_{\text{ad}} = \frac{a}{U_{\text{max}}}. \tag{8}$$

## 2.2 Numerical Model

The numerical model consists of a partitioned fluid-structure interaction solver in which the lattice Boltzmann (LB) method is used for the liquid, the finite element (FE) method for the capsule dynamics, and the immersed boundary (IB) method for the fluid-structure interaction. This IB-LB-FE solver has previously been employed in the study of the dynamics of deformable red blood cells and capsules (Krüger *et al.*, 2013, 2014). Here, we provide essential properties of the model, while comprehensive details are available elsewhere (Krüger *et al.*, 2011).

For the LB method, we use the D3Q19 lattice (Qian *et al.*, 1992) and the BGK collision operator (Bhatnagar *et al.*, 1954) with relaxation time $\tau$. The viscosity of the liquid and the relaxation time satisfy

$$\nu = c_s^2 \left( \tau - \frac{\Delta t}{2} \right) \tag{9}$$

where $c_s$ is the lattice speed of sound and $\Delta t$ is the time step. For the D3Q19 lattice, $c_s^2 = \Delta x^2/(3\Delta t^2)$ holds where $\Delta x$ is the lattice resolution. For the channel flow cases, flow is driven by a constant body force following the forcing method of Guo *et al.* (2002). This form of the LB method is widely used in the field of fluid dynamics, including in previous inertial microfluidics studies (Schaaf & Stark, 2017; Schaaf *et al.*, 2019).

Each capsule is represented by a surface mesh with $N_f$ flat triangular faces (or elements) defined by three nodes (or vertices) each. At a given time step, the capsule mesh is generally deformed and the deformation state of each face (for shear and area deformation) and the angles between pairs of faces (for bending deformation) are used to calculate the force on each vertex. To prevent contact between capsules, a repulsion force is used between pairs of nodes belonging to different capsule meshes when both nodes come closer than $\Delta x$ (Krüger *et al.*, 2013).

We employ an IB method with a 3-point stencil to interpolate velocities and spread forces between the Eulerian (fluid) and Lagrangian (membrane) meshes (Peskin, 2002). This treatment recovers the no-slip boundary condition at the surface of the capsules and the momentum exchange between the liquid and the capsule membrane.

The no-slip boundary condition at the resting and moving walls for channel and shear flow, respectively, is realised by the standard half-way bounce-back condition (Ladd, 1994). The flow is periodic along those directions that are not confined by walls (flow direction in the channel case, and shearing plane in the shear flow case). The channel length $L$ is chosen sufficiently long to avoid the interaction of capsules with their periodic images.

## 3 Benchmark Tests

We test our model by comparing simulation results obtained from our solver with previously published results from other groups: trajectories of a soft particle pair in shear flow (section 3.1) and lateral migration of a single soft capsule in channel flow (section 3.2).

### 3.1 Soft Particle Pair in Shear Flow

Doddi & Bagchi (2008) numerically investigated the effect of inertia on the interaction between a soft particle pair in a shear flow, whereby the particles are slightly offset from the centre of the channel as shown in Fig. 1. By increasing the shear rate $\dot{\gamma}$, the trajectories of the pair switch from passing to reversing. Doddi and Bagchi employed the neo-Hookean model, Eq. (2), for the capsules and the front-tracking/immersed boundary method proposed by Unverdi & Tryggvason (1992). Here, we reproduce these results with our solver to test the accuracy of the IB-LB-FE method and its ability to capture hydrodynamic interactions of soft capsules in mildly inertial flows. Tab. 1 shows the relevant parameters of this benchmark case.

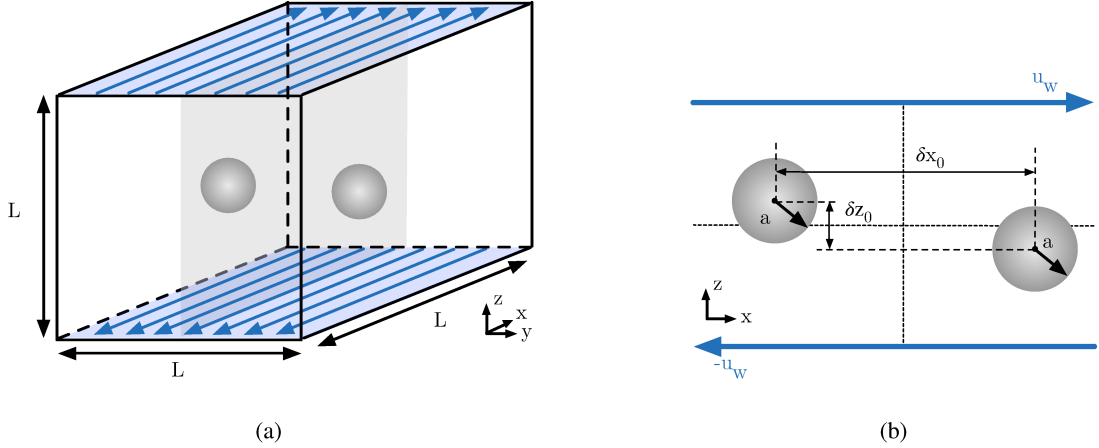

(a)                                                                               (b)

Figure 1: Schematic of the first benchmark case: soft particle pair in shear flow. (a) Both particles are located on a plane parallel to the shear direction. (b) The shear rate is defined by the speed of the moving walls and their separation: $\dot{\gamma} = 2u_\mathrm{w}/L$. Both particles have the same distance from the walls, where one particle is closer to the bottom and the other closer to the top wall. Particles are initially separated by $\delta x_0$ and $\delta z_0$ along the $x$- and $z$-axes, respectively. Simulation parameters are reported in Tab. 1.

While Doddi & Bagchi (2008) did not report whether they used a bending resistance, the work originally employing the numerical method included a finite bending resistance to avoid folding of the membrane (Eggleton & Popel, 1998). Therefore, we ran two sets of simulations: one without bending resistance, the other with a finite bending resistance ($\tilde{\kappa}_\mathrm{b} = 0.00287$). Fig. 2(a) shows our simulated particle trajectories for different values of $\mathrm{Re}_\mathrm{p}$ in the plane highlighted in Fig. 1 in comparison with the results of Doddi & Bagchi (2008). Good agreement is seen between the original results and our simulations that include a bending resistance, while simulations without bending resistance show some differences in the later stages of the passing trajectories. Independent of the bending resistance, we see that, as $\mathrm{Re}_\mathrm{p}$ is increased, each particle moves closer to the midplane between the walls, resulting in the transition from passing to reversing trajectories. Using the IB-LB-FE model, we observed this transition between $\mathrm{Re}_\mathrm{p} = 0.375$ and $0.575$, in agreement with previous results.

Doddi & Bagchi (2008) employed the neo-Hookean model only. Since other elastic capsule models are often used, we have repeated the same test for capsules equipped with the Skalak model to investigate the sensitivity of the trajectories to the details of the capsule model. The reduced bending modulus is $\tilde{\kappa}_b = 0.00287$ for both elastic models, and $\tilde{\kappa}_\alpha = 2$ for the Skalak model (note that $\tilde{\kappa}_\alpha$ is not defined for the neo-Hooken model). Fig. 2(b) shows the comparison of trajectories for the neo-Hookean and the Skalak models as obtained from the IB-LB-FE solver. Overall, details of the trajectories are not significantly altered by the model. The Skalak model leads to slightly

| Parameter | Value |
|---|---|
| $\mathrm{Re}_\mathrm{p}$ | 0.125, 0.375, 0.575, 0.75 |
| Ca | 0.025 |
| $\nu$ | $1/6 \Delta x^2/\Delta t$ |
| $\kappa_\mathrm{b}$ | 0, 0.00287 |
| $a$ | $14.4 \Delta x$ |
| $L$ | $12.5a = 180 \Delta x$ |
| $\delta x_0$ | $4.0a = 57.6 \Delta x$ |
| $\delta z_0$ | $0.2a = 2.88 \Delta x$ |

Table 1: Parameters of the first benchmark case: soft particle pair in shear flow. See Fig. 1 for an illustration of the setup. The shear rate $\dot{\gamma}$ depends on $\mathrm{Re}_\mathrm{p}$ according to Eq. (4), and the shear elasticity $\kappa_\mathrm{s}$ is obtained from Eq. (6). The liquid density is set to 1 in simulation units.

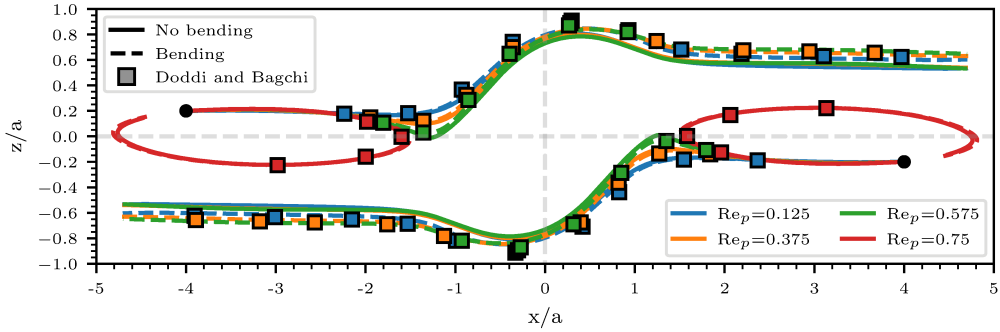

(a) Comparison of our simulations with original results.

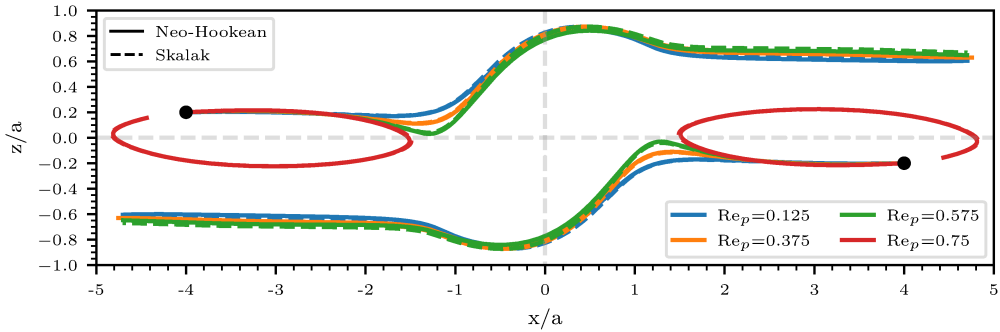

(b) Comparison between neo-Hookean and Skalak models.

Figure 2: Trajectories of particle pairs in simple shear flow for various Reynolds numbers. Blue: $\text{Re}_p = 0.125$; orange: $\text{Re}_p = 0.375$; green: $\text{Re}_p = 0.575$; red: $\text{Re}_p = 0.75$. Black circles indicate the initial position of each particle. See Fig. 1 for the geometry setup and Tab. 1 for simulation parameters. (a) Comparison of our results for the neo-Hookean model with and without bending resistance with previous data (Doddi & Bagchi, 2008). Squares are data points extracted from Fig. 8 in Doddi & Bagchi (2008) using WebPlotDigitizer v4.4. (b) Comparison between neo-Hookean and Skalak models with bending resistance obtained from our IB-LB-FE solver.

larger lateral displacements. This difference is probably caused by the reduced deformation of the capsules due to the strain-hardening properties of the Skalak model compared to the strain-softening neo-Hookean model. While the difference between the constitutive models is not trivial, the trajectories are qualitatively similar, and the transition between passing and reversing trajectories is also between $\text{Re}_p = 0.375$ and $0.575$ for the Skalak model.

## 3.2 Lateral Migration of Single Soft Particle in Channel Flow

Before investigating the interaction of a particle pair in channel flow with moderate inertia, it is crucial to ensure the inertial effects on a single particle are captured accurately. We consider the inertial migration of a single soft capsule in a pressure-driven Poiseuille flow through a square duct with edge length $2w$ and axial length $L = 8w$ as shown in Fig. 3. Periodic boundary conditions are used along the flow axis.

Schaaf & Stark (2017) investigated the effect of Laplace number on the lateral equilibrium position of a soft capsule at a given channel Reynolds number in the same geometry. They demonstrated that the lateral equilibrium position moves away from the channel centre as La increases. For the case $\text{Re}_c = 100$, a transition occurs where the equilibrium position switches from a diagonal location to a channel face centre. As a hallmark of lateral migration in a square duct, one can generally distinguish two phases of migration: a first phase during which the capsule quickly migrates in radial direction, and a second phase defined by a slower circumferential migration along a heteroclinic orbit (Nakagawa *et al.*, 2015). This behaviour is caused by the shear gradients of the velocity field being more pronounced in radial than in circumferential direction.

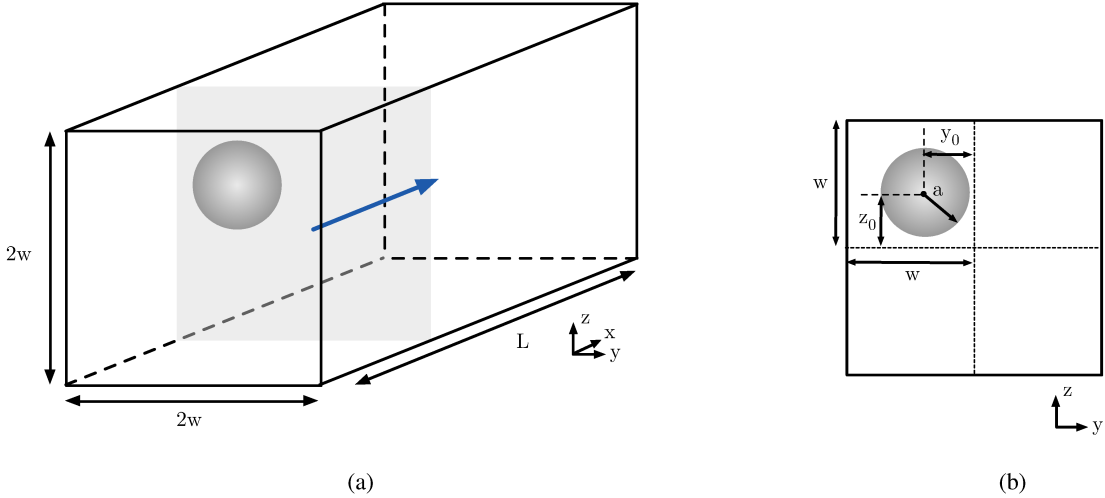

(a)                  (b)

Figure 3: Schematic of the first benchmark case: soft particle migration in channel flow. (a) The arrow indicates the flow direction. The grey plane indicates the channel cross section. (b) The particle is initially located away from the channel centreline. Simulation parameters are reported in Tab. 2.

We have run simulations for $\mathrm{Re_c}$ = 10 and 100, and for La between 1 and 100. The Skalak model was employed for the capsule membrane. Tab. 2 summarises the simulation parameters. The initial position of the capsule is the same for each set of parameters considered. We compare cross-sectional particle trajectories obtained from our solver with those results reported by Schaaf & Stark (2017). Fig. 4 shows that the capsule trajectories and lateral equilibrium positions generally agree well. In the cases where results do not agree well, the equilibrium positions recovered by our solver are located either on the channel diagonals or midway along channel faces, as expected from symmetry considerations, whereas the trajectories of Schaaf *et al.* appear to have stopped short. This discrepancy could be explained by the runtime of the simulations by Schaaf & Stark (2017) who mentioned that not all of the particles "reach their equilibrium position on the diagonal or the main axis during the simulations".

We also noticed some oscillations in the results of Schaaf *et al.* at La = 5 and 10 that appear to affect the trajectories during the second phase of migration. Our simulations did not show these oscillations, which could be due to a different bending model employed.

Overall, our IB-LB-FE solver produces results consistent with those of previously validated solvers and has been shown to be suitable for the investigation of inertial migration and the hydrodynamic interaction of soft capsules.

| Parameter | Value |
|---|---|
| $\mathrm{Re_c}$ | 10, 100 |
| La | 1, 5, 10, 50, 100 |
| $\nu$ | $1/6 \Delta x^2/\Delta t$ |
| $\tilde{\kappa}_\alpha$ | 2 |
| $\tilde{\kappa}_b$ | 0.00287 |
| $a$ | $9\Delta x$ |
| $w$ | $5a = 45\Delta x$ |
| $L$ | $8w = 360\Delta x$ |

Table 2: Parameters of the second benchmark case: soft particle migration in channel flow. See Fig. 3 for an illustration of the setup. The channel Reynolds number is varied by the body force and therefore $U_{\max}$ via Eq. (5), and the Laplace number is controlled by the shear elasticity via Eq. (7). The liquid density is set to 1 in simulation units.

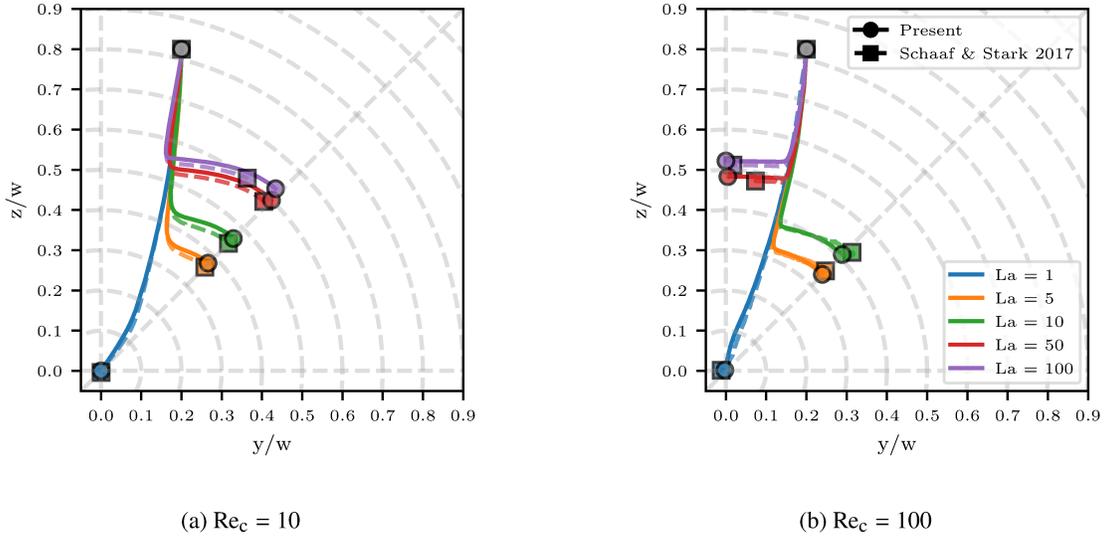

(a) $\text{Re}_c = 10$

(b) $\text{Re}_c = 100$

Figure 4: Cross-sectional trajectories and lateral equilibrium positions of a single capsule in channel flow for various values of La at (a) $\text{Re}_c = 10$ and (b) $\text{Re}_c = 100$. Our results are indicated by solid lines with circles, and dashed lines with squares show the results reported by Schaaf & Stark (2017). Grey symbols mark the initial capsule position, and colourful symbols indicate the final positions on the channel cross section. The channel centreline is located at $(y, z) = (0, 0)$. Dashed grey lines are guides for the eyes.

## 4 Results and Discussion

Previous works have investigated the formation and stability of pairs (Gupta *et al.*, 2018; Udono, 2020) and trains (Hu *et al.*, 2020; Kahkeshani *et al.*, 2016) of rigid particles in channel flow. Schaaf *et al.* (2019) identified four different types of the trajectories of two rigid particles, depending on their initial positions in the channel. Patel & Stark (2021) also observed each of these trajectory types for soft particles in mono- and bidisperse pairs. In the following, we analyse the interaction of a pair of equally soft capsules for different Laplace numbers. In section 4.1 we define the cases investigated and parameters used, while section 4.2 presents the trajectory types that occur at a given Laplace number. Section 4.3 investigates the relationships of the lateral equilibrium position of each particle and the axial equilibrium distance between both particles with Laplace number and initial positions. Section 4.4 analyses the formation of stable pairs with a view to understanding how Laplace number and initial position effect focusing time and distance.

### 4.1 Case Definition

Originally proposed by Schaaf *et al.* (2019), two particles are placed in a pressure-driven flow through a rectangular duct with width $2w$ and height $2h$ with aspect ratio $w/h = 2$ (Fig. 5). The length of the channel is $L = 10h$, and the flow is periodic along the flow direction.

Both capsules are initially located on the mid-plane between the side walls ($y = \text{const}$), while initial $x$- and $z$-coordinates are varied. We distinguish between the initially leading (farther downstream) and lagging (farther upstream) particles, according to their initial positions on the flow axis ($x$-axis). The limitation to the mid-plane is justified since particle equilibrium positions in high aspect-ratio channels at the Reynolds number investigated here are usually in the middle of the long channel edges (Prohm & Stark, 2014). Particles initially located on this mid-plane will usually stay on this plane while moving along the $x$-axis and migrating along the $z$-axis. We have not observed particles leaving the mid-plane in any of our simulations. We expect that changing the channel aspect ratio $w/h$ will have only a minor influence on the results as long as the confinement by the side walls is lower than that by the top and bottom walls. Investigating the more general cases with arbitrary channel aspect ratios and initial particle positions not confined to the mid-plane would lead to unmanageable number of free parameters and is beyond the scope of this paper.

In all following simulations, we have employed the Skalak membrane model with a reduced dilation modulus $\tilde{\kappa}_\alpha = 2$ and reduced bending modulus $\tilde{\kappa}_b = 0.00287$. The channel Reynolds number is $\text{Re}_c = 10$ in all cases, and the numerical viscosity is kept at $\nu = 1/6 \, \Delta x^2 / \Delta t$. The channel dimensions are $2w = 80\Delta x$, $2h = 160\Delta x$ and $L = 560\Delta x$. The undeformed capsule radius is $a = 16\Delta x$. We focus on La, $\delta x_0$, $z_0^{\text{lead}}$ and $z_0^{\text{lag}}$ as free parameters (Fig. 5).

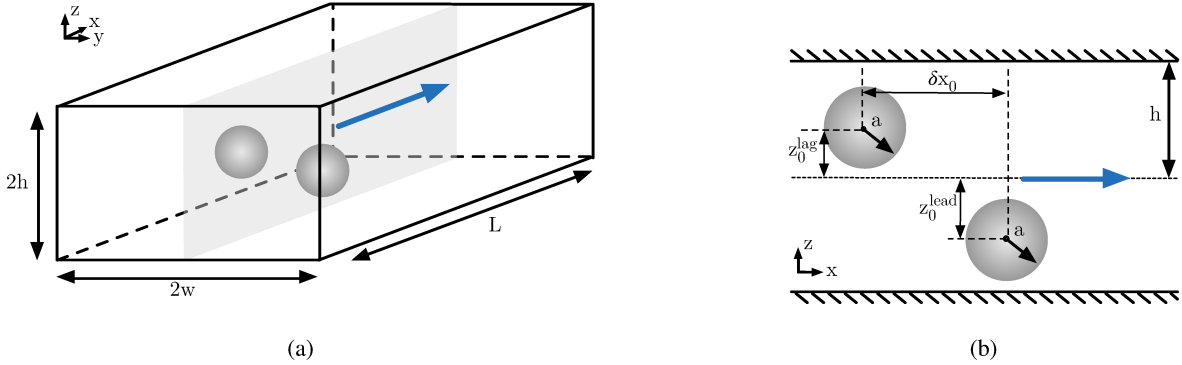

(a)                                                         (b)

Figure 5: Schematic of particle pairs in a rectangular duct. Parameter values are given in section 4.1. (a) The flow is along the $x$-axis (blue arrow). Particles are initially located on the mid-plane with $y$ = const (indicated by grey plane). (b) Depending on their initial position along on the $x$-axis, we distinguish between the leading and the lagging particle.

## 4.2 Interaction Types of Homogeneous Soft Particle Pairs

We consider pairs of identical capsules (homogeneous pairs). First, we characterise the interaction types of the capsules before investigating the effects of Laplace number and initial particle positions in more detail.

In order to understand the general effect of softness and initial positions of the capsules, we simulated pairs at three different Laplace numbers: La = 1, 10 and 100. The initial positions of both capsules along the height axis ($z$-axis) is varied with a total of 66 combinations (six initial positions of the leading particle: five in the top half of the channel and one at the centre; 11 initial positions of the lagging particle: five in the top half, five in the bottom half and one at the centre). Note that initial configurations mirrored at the $x$-$y$-plane lead to identical results. We also considered three different initial axial distances between the particle centres ($3a$, $5a$ and $7a$), leading to $3 \times 66 \times 3 = 594$ configurations in total. The actual number of simulations is 549 since there are 45 symmetric cases when the leading particle is initially midway between the bottom and top walls.

Using rigid particles, Schaaf *et al.* (2019) identified four different particle interaction types: *Swap & Scatter*, *Pass & Scatter*, *Scatter* and *Capture*. Examples of each trajectory type are shown in Fig. 6, and brief descriptions are included in Table 3. We changed the original names of the trajectory types that Schaaf *et al.* (2019) used in order to improve clarity when discussing the trajectories observed for soft particles. 'Scatter' refers to cases where the axial distance between particles eventually grows such that particles stop interacting with each other. 'Capture' indicates an interaction where the axial distance of the two particles is bound. Interestingly, we observe two capture sub-types: one where the axial (and also lateral) distance between the particles becomes constant after some time ('stable') and the other where the axial distance does not grow arbitrarily, but does not converge to a constant value either ('partially stable').

Using soft particles, we identified the same four previously observed trajectory types of rigid particles, and two new types that have not been found for rigid particles. We label the two new types as *Swap & Capture* and *Pass & Capture*. Examples of the new types are shown in Fig. 6 and described in Table 3. The newly observed partially stable particle pairs seem to occur only when particles are sufficiently soft, as will be discussed later.

Fig. 7 shows the key results of our work: trajectory types resulting from all studied configurations (various initial lateral positions; initial axial particle distances of $3a$, $5a$ and $7a$; Laplace numbers La = 1, 10 and 100). On first inspection, a general trend can be seen that particle capture occurs under a wider range of initial conditions when the particles are softer (smaller La). Furthermore, particle capture is more likely when the initial axial distance $\delta x_0$ is smaller. For example, for the extreme case La = 1 and $\delta x_0 = 3a$, 90% of all studied initial positions lead to capture. Only in cases where the leading particle is initially much closer to the channel centre (and therefore sufficiently faster) than the lagging particle, capturing does not occur.

The increase of capture probability with decreasing $\delta x_0$ can be explained by the stronger hydrodynamic interaction of particles when they are initially closer. Additionally, particle softness is beneficial for capture since an initially lagging particle that is softer is able to increase its axial velocity more quickly than a more rigid particle by migrating closer towards the channel centre where the axial free-stream flow velocity is higher. Thus, a softer lagging particle has a higher chance of catching up with the leading particle before the leading particle moves away. We will take a closer look at the effect of Laplace number on particle dynamics in section 4.3.

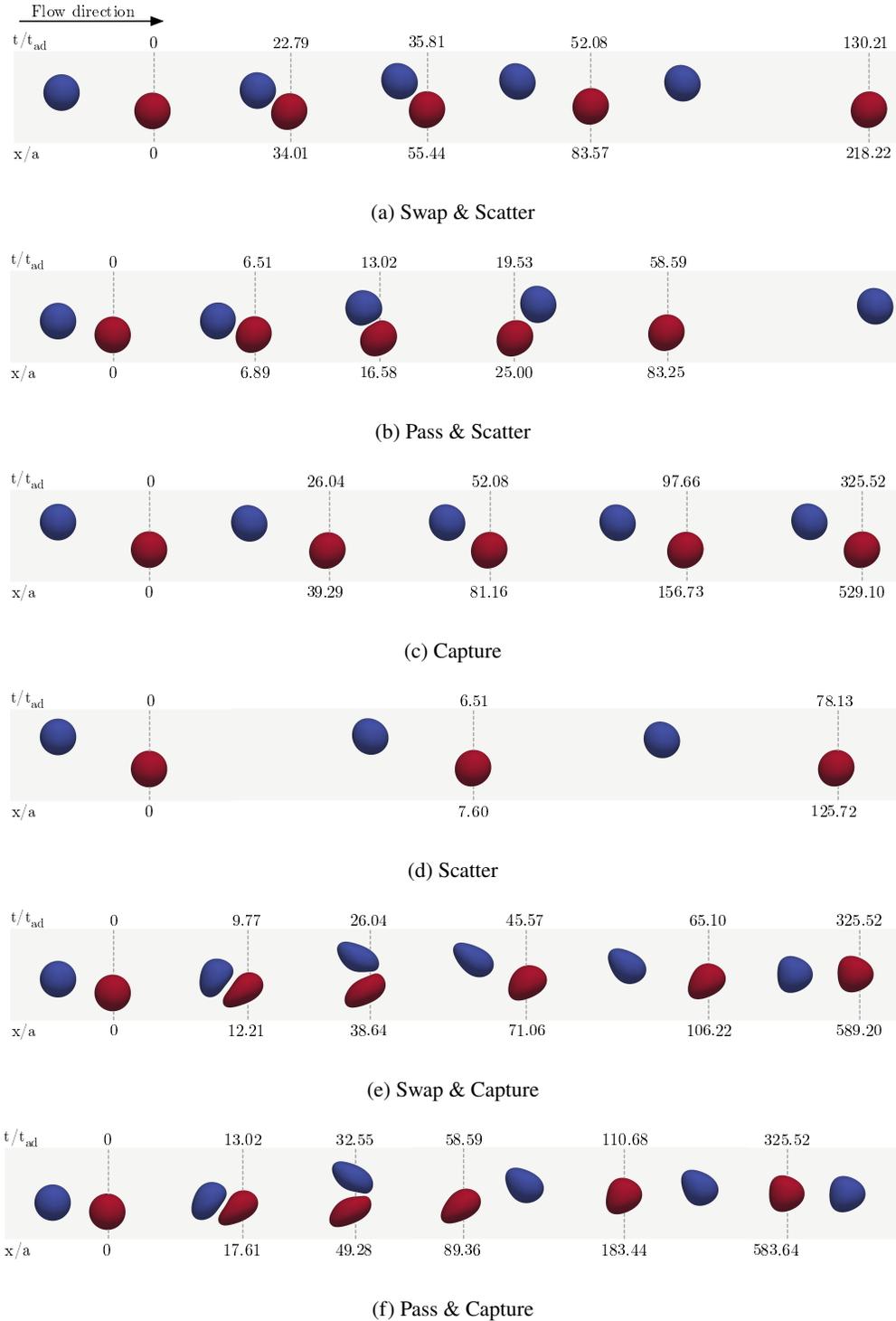

Figure 6: Typical snapshots at selected points in time and space for each trajectory type observed for soft particles. Cases (a)–(d) have been previously observed for rigid particles as well, while cases (e)–(f) have not. The initially leading/lagging particle is shown in red/blue, respectively. Note that the axial distances between different snapshots are not to scale. The six interaction types are characterised in more detail in Tab. 3.

| Type | Description | Rigid |
|---|---|---|
| (a) Swap & Scatter | The lagging particle approaches the leading particle and attempts to overtake. As it does so, the particles swap their lateral positions, which means that the particle that was initially farther away from the channel centre is now closer, and vice versa. The lagging particle fails to overtake the leading particle and, due to the swapping of the lateral positions, is now farther away from the channel centre than the leading particle. The leading particle is now faster and moves away from the lagging particle. | Yes |
| (b) Pass & Scatter | The lagging particle approaches the leading particle and overtakes on the side closer to the channel centre. The initially lagging particle (now leading particle) is closer to the channel centre and moves away from the initially leading (now lagging) particle. | Yes |
| (c) Capture | The lagging particle approaches the leading particle but does not overtake. Instead, the lagging particle follows the leading particle at a constant distance. In some cases, a damped oscillation of the axial distance between the particles occurs. | Yes |
| (d) Scatter | The leading particle is faster than the lagging particle and moves away. | Yes |
| (e) Swap & Capture | Similarly to the Swap & Scatter trajectory, the lagging particle approaches the leading particle and the lateral positions swap. However, once the lagging particle fails to overtake, it follows the leading particle, not necessarily at a constant distance. | No |
| (f) Pass & Capture | Similarly to the Pass & Scatter trajectory, the lagging particle overtakes the leading particle on the side closest to the channel centre. Once it has overtaken, the initially lagging particle begins to move away. However, the now lagging particle is able to follow, not necessarily at a constant distance. | No |

Table 3: Descriptions of trajectory types of soft particle pairs with indication of whether this type is also observed for rigid particle pairs. Scatter means that the axial distance between particles grows until particles stop interacting with each other. Capture means that the axial distance is bound. Fig. 6 visualises some example cases.

Another notable effect of particle softness is that the diagrams in Fig. 7 become more symmetric with respect to $z_0 \leftrightarrow -z_0$ of the lagging particle. For La = 100 (more rigid particles), capture essentially only occurs when the leading and lagging particles are on different sides of the channel (positive $z_0$ for leading particle and negative $z_0$ for lagging particle). This effect can be explained by the known observation that pairs of rigid particles at high confinement $a/h$ are more stable when both particles are located on different sides of the channel centre (Patel & Stark, 2021). Softer particles, however, migrate more quickly, and they migrate closer to the channel centre; therefore, the initial lateral position of softer particles is less important. Furthermore, the faster lateral migration of softer particles provides more opportunity for a soft lagging particle to remain within interaction range of the leading particle, resulting in a captured pair rather than being scattered.

Schaaf *et al.* (2019) did not observe the *Swap & Capture* and *Pass & Capture* types for rigid particle pairs. Fig. 7 shows that both types become less common when La increases. Thus, the results suggest that a critical Laplace number exists where these trajectory types disappear. Similarly, we found partially stable particle pairs only for La < 100.

Examination of Fig. 7 reveals that a small variation in the initial position can have a large effect on the trajectory type of the two particles. To investigate this effect further, two sets of configurations are selected where the Laplace number, the initial axial distance between the particles and the initial lateral position of the leading particle remain constant. The initial lateral position of the lagging particle is varied between $-0.5h$ and $+0.5h$. The selection of initial conditions is highlighted by the dashed areas in Fig. 7.

The time evolution of the axial distance between both particles is shown in Fig. 8. For a given value of La and changing initial position of the lagging particle, various trajectory types can be seen, with stable pairs forming in some cases. Softer particle pairs (La = 10) result in more trajectory types where captured particle pairs are created, but some similarities exist between the cases for La = 10 and 100. For La = 10, the trajectory type transitions from *Capture* when the lagging particle is located at the extreme of the opposite side of the channel centreline ($z_0^{\text{lag}} = -0.5h$) to *Swap & Capture* as the lagging particle is released closer to the centreline. Similarly, for La = 100, the *Capture* trajectory type also occurs when the lagging particle is at the extreme of the opposite side of the channel centreline. Once the initial position of the lagging particle is closer to the channel centreline, the early interaction for the stiffer particles (La = 100) is similar to that of the softer particles (La = 10) with some degree of swapping of lateral positions between the particles. However, differences begin to occur after this early interaction: the lagging particle moves away from the leading particle for La = 100, resulting in a *Swap & Scatter* type. Upon $z_0^{\text{lag}}$ becoming positive (*i.e.* both particles are initially on the same side of the channel), another trajectory type transition occurs: for La = 10, we find a transition to *Pass & Capture*, while the pair at La = 100 transitions to *Pass & Scatter*, marking a further difference between the cases for both Laplace numbers. The point of transition to a passing trajectory is also La-dependent; $z_0^{\text{lag}} = 0$ for La = 100 and $z_0^{\text{lag}} = 0.2h$ for La = 10. Finally, for the largest studied values of $z_0^{\text{lag}}$, the trajectory types transition to *Scatter*, irrespective of the Laplace number. All transitions described here are also visible in Fig. 7.

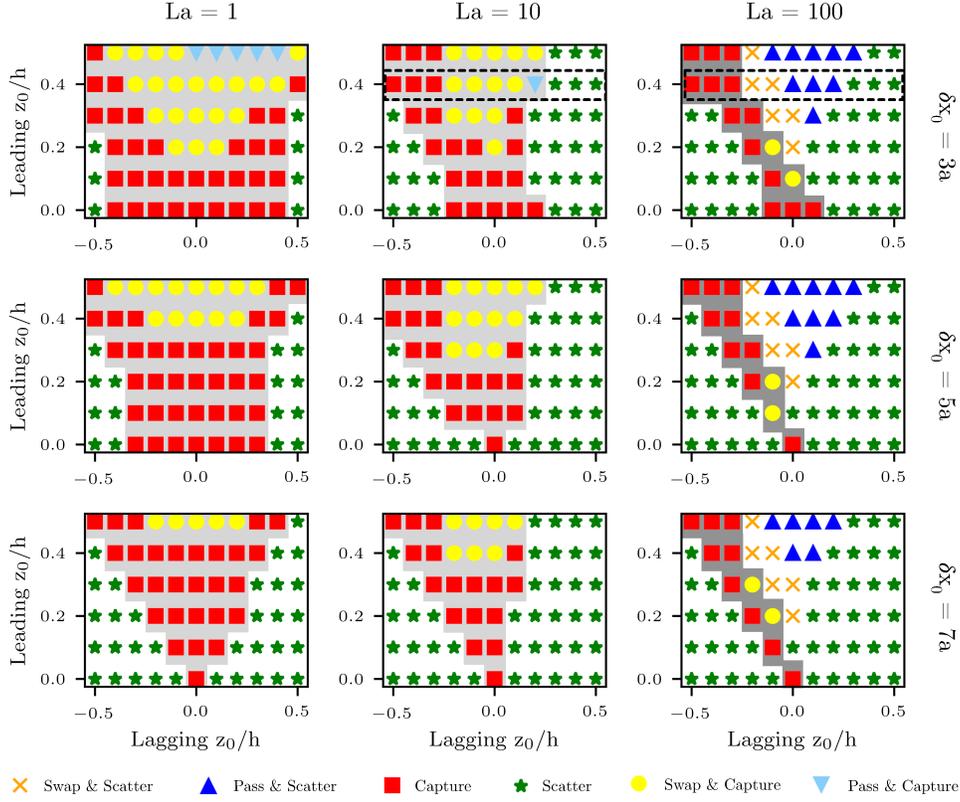

Figure 7: Particle interaction types as function of Laplace number, initial axial distance $\delta x_0$ and initial lateral positions $z_0$ for homogeneous pairs. The *x*- and *y*-axes of each panel indicate the lateral initial positions of the lagging and leading particles with respect to the channel centreline ($z_0 = 0$). The symbols indicate the interaction type as defined in Tab. 3 and shown in Fig. 6. Symbols with a white background indicate a scattering trajectory. A dark grey background denotes a stable pair, while a light grey background indicates a partially stable pair. The dashed boxes define those configurations that are analysed in more detail in Fig. 8.

On a finer sweep of the initial position range of the lagging particle where $z_0^{\text{lag}}$ was varied by increments of $0.033h$ (data not shown), no additional transitions were found, confirming that not all trajectory types exist for a given Laplace number.

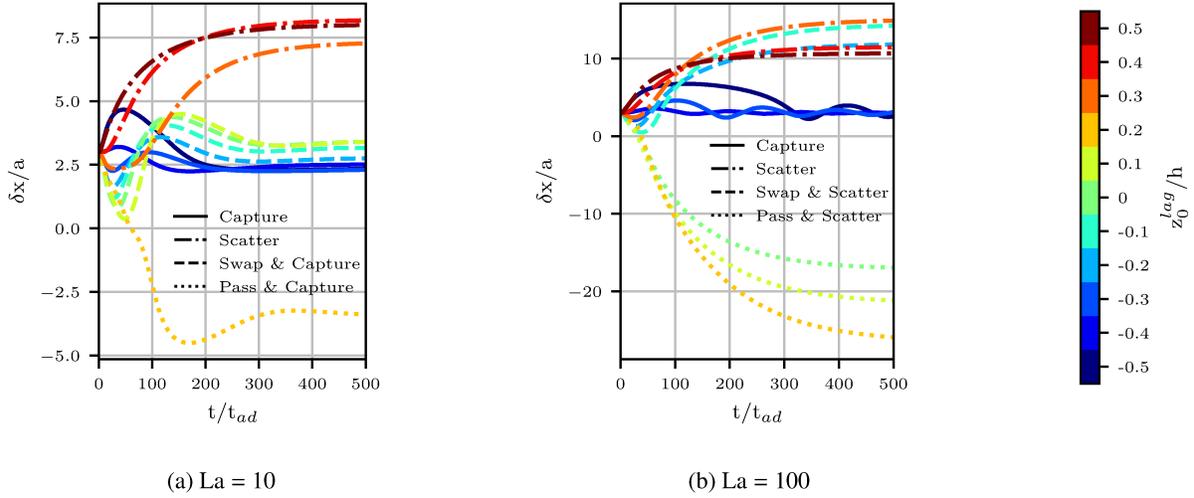

(a) La = 10

(b) La = 100

Figure 8: Time evolution of the axial distance $\delta x$ for various initial lateral positions of the lagging particle (shown in different colours) for (a) La = 10 and (b) La = 100. The line style denotes the resulting trajectory type. In all cases, $z_0^{\text{lead}} = 0.2h$ and $\delta x_0 = 3a$.

For the same system properties (Reynolds number, Laplace number, channel aspect ratio, particle-channel confinement), different initial particle configurations can lead to entirely different outcomes: captured or scattered pairs. Captured pairs are observed within a wider range of initial positions when the particles are softer. Furthermore, depending on the initial configuration, the axial distance between particles in (partially) stable pairs can be different. Fig. 8(a) shows the partially stable axial distance can vary between around $2.5a$ and $3.5a$, a variation of nearly 50%. For the *Swap & Capture* and *Pass & Capture* types, the axial distances are different and also vary with the initial particle position. For the *Capture* trajectory type at La = 10, the variation in final axial distance is smaller, however remains dependent on the initial particle position. In contrast, the final axial distance of *Capture* trajectory type pairs at La = 100 are independent of initial positions (Fig. 8(b)). We explore this behaviour further in section 4.3 since these are important observations given the formation of stable pairs is useful in many microfluidic applications when the particle spacing within a channel is required to be predictable and reproducible.

As we will elaborate next, the role of Laplace number and lateral equilibrium position are tightly connected with the stability of a captured particle pair.

## 4.3 Effect of Softness on Lateral and Axial Particle Migration

After having identified different interaction types of soft particle pairs, we now turn our attention to the dynamics of particle capture. In their investigation of rigid particle pairs under the same flow conditions as this work, Schaaf *et al.* (2019) found that stable particle pairs form for certain initial lateral positions. Rigid particles show a damped oscillation of their relative distance before reaching their stable equilibrium configuration. We observed a similar oscillation for soft particles.

We investigated the oscillation of the relative distance between soft particles for $z_0^{\text{lead}} = 0.2h$, $z_0^{\text{lag}} = -0.2h$ and $\delta x_0 = 3a$ in the range La = [5, 125]. All these cases have *Capture* trajectories and lead to either stable or partially stable pairs. Fig. 9(a) shows the time evolution of the lateral position of both particles as function of La. In the early stages of migration, each particle undergoes a damped oscillation before reaching a La-dependent lateral equilibrium position. Trajectories of single particles at the same La are included for comparison. Single particles follow similar migration paths to the particles within the pair of equal stiffness but without oscillation, demonstrating that the oscillations are a result of particle-particle interaction. While the softest particles ($La \leq 15$) migrate to the channel centre, stiffer particles reach equilibrium positions farther away from the channel centre as La increases, converging to rigid limits extracted from Schaaf *et al.* (2019). This observation is in agreement with the results of Kilimnik *et al.* (2011).

Fig. 9(b) shows a zoomed area of Fig. 9(a), highlighting the transition between off-centre and centreline lateral equilibrium positions, also in comparison with the single particle lateral trajectories. The point of transition between

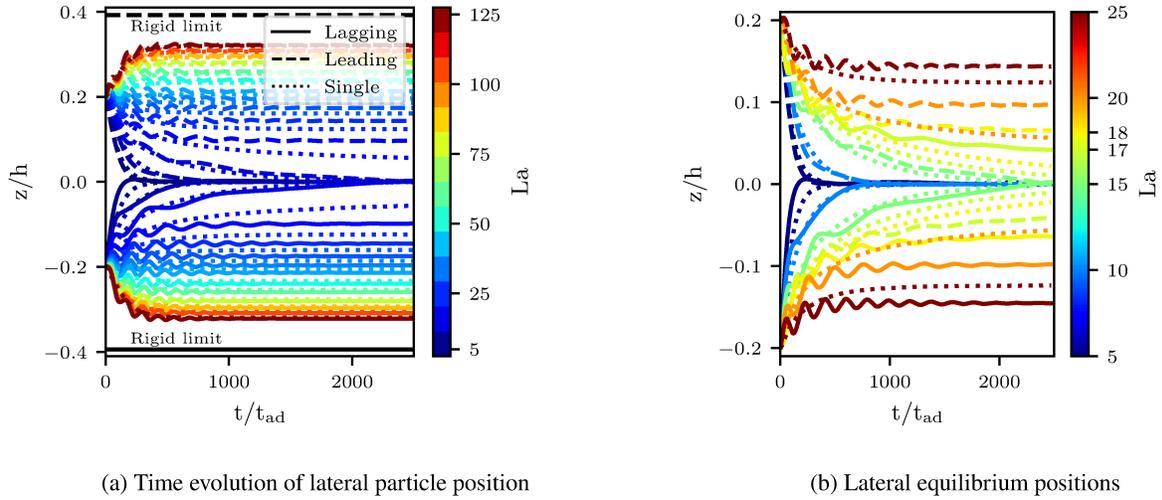

(a) Time evolution of lateral particle position

(b) Lateral equilibrium positions

Figure 9: Lateral motion of soft particles for $\delta x_0 = 3a$ and various Laplace numbers. (a) Time evolution of single, leading and lagging particles. Rigid limits extracted from Schaaf *et al.* (2019) for lateral equilibrium positions of rigid particles under the same flow conditions. (b) Zoomed area of (a) in the region of transition between off-centre and centreline lateral equilibrium positions.

off-centre and centreline equilibrium positions for a single particle is termed the critical Laplace number and was found to be $La_{cr} \approx 18$ for a single particle (the full migration time at the critical Laplace number is large and not fully shown in Fig. 9(b)). However, Fig. 9(b) shows a particle pair with $La = 18$ migrating to an off-centre equilibrium position, as does a slightly softer pair with $La = 17$. These observations demonstrate that the formation of pairs slightly alters the inertial migration characteristics of soft particles.

The damped oscillation is also seen in the axial distance between both particles (Fig. 10). While particle pairs with $La > 15$ converge to a stable axial equilibrium distance, softer particles fail to do so. However, their axial distance remains small during the runtime of the simulations, indicating that these pairs are partially stable, rather than scattered. We find that pairs that form stable axial equilibrium distances have off-centre lateral equilibrium positions, while partially stable pairs have centreline lateral equilibrium positions in all cases. This observation can be attributed to the absence of shear-gradient-induced lift at the channel centreline. As a result, other forces have larger contributions to the overall dynamics of the system while the influence of numerical artefacts cannot be fully ruled out.

We observed that, for the investigated parameter range, all pairs at the centreline are partially stable while all stable pairs are off-centre. We hypothesise that the determining factor for stability is lateral equilibrium position, rather than particle softness (Laplace number). For a given confinement and Reynolds number, the Laplace number determines the lateral equilibrium position. In particular, at $\chi = 0.4$ and $Re = 10$, we found $La_{cr} \approx 15$. To test the hypothesis, we investigated the behaviour of a particle pair at $\chi = 0.2$ and $La = 15$ for which we expect a single particle and particle pairs to assume an off-centre equilibrium position. Fig. 11 shows the trajectories of particle pairs for two initial axial distances: $\delta x_0 = 3a$ to have the same $\delta x_0/a$ ratio and $\delta x_0 = 6a$ to have the same $\delta x_0/h$ ratio as the case with $\chi = 0.4$. The lateral migration of a single particle with same softness, confinement and initial lateral position is included for comparison. Both sets of initial conditions lead to the pair migrating to an off-centre equilibrium position (Fig. 11(a)) and to stable pairs forming (Fig. 11(b)). Hence, the findings for $\chi = 0.2$ provide evidence that stable pairs require off-centre equilibrium positions. Applications relying on finely tuned axial distances between particles, therefore, might benefit from arrangements with sufficiently small $\chi$ and large $La$ for which particles do not migrate to the centreline.

The dependence of axial distance on initial configuration when particle are located at the centreline may explain some observations previously reported where the axial distance between particles is distributed within a certain range, with most of the measured distances clustered around one or several preferred distances (Lee *et al.*, 2010; Kahkeshani *et al.*, 2016). In realistic problems, a channel contains more than two particles, and each *Scatter* interaction means that the leading particle may catch up with another particle that is farther downstream in the channel. This way, after some time, stable trains might form, even if individual pairs are not stable. Future studies should therefore investigate how

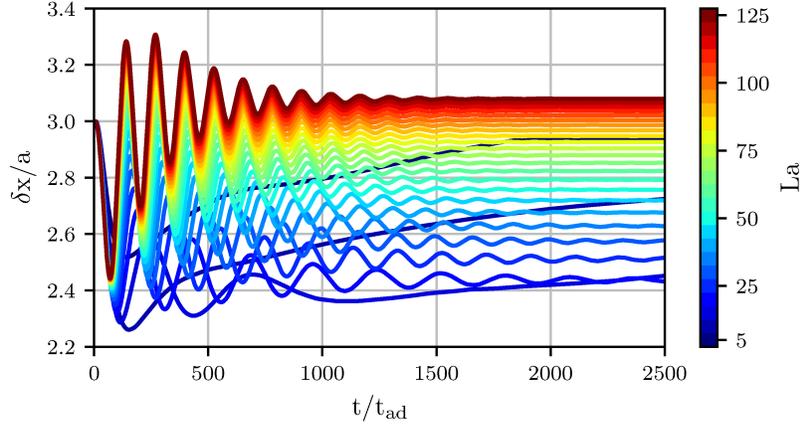

Figure 10: Time evolution of axial distance for various Laplace numbers. The initial conditions are the same as in Fig. 9.

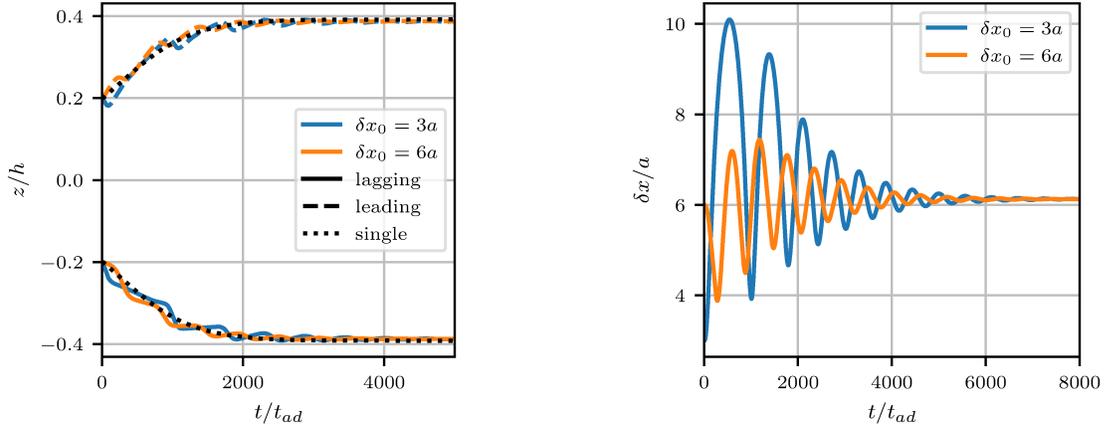

(a) Time evolution of lateral particle positions.

(b) Time evolution of axial particle distance.

Figure 11: Time evolution of single particle and particle pair motion with particles of La = 15 and $\chi$ = 0.2. Initial lateral positions of the particles are identical to Fig. 9 and Fig. 10 in non-dimensional units while the initial axial distance is equal to Fig. 9 and Fig. 10 in terms of channel height ($\delta x_0 = 6a$) and in terms of particle radius ($\delta x_0 = 3a$).

the dynamics of single pairs is related to the dynamics of longer particle trains.

Fig. 12(a) shows the absolute value of the lateral equilibrium position of the leading and lagging particles as function of inverse La; the equilibrium position of a single particle under the same conditions is included for comparison. Also included for comparison is the equilibrium position of a rigid pair as reported by Schaaf *et al.* (2019) with which we obtain good agreement. There are two important findings beside the fact that softer particles end up closer to or at the channel centre. First, the absolute values of the lateral positions of both particles in a pair are virtually indistinguishable from each other. However, the lagging particle is slightly farther from the channel centre than the leading particle. This finding was also observed by Schaaf *et al.* (2019) for rigid particles. Second, both particles in a pair are consistently farther away from the channel centre than a single particle under the same conditions. This difference becomes smaller for increasing La, and our findings suggest that in the rigid limit ($1/\text{La} \to 0$), lateral equilibrium positions are the same for a single particle and particles in a pair. Indeed, Schaaf *et al.* (2019) found rigid particles having the same lateral equilibrium position when in a pair and in isolation. Interestingly, Gupta *et al.* (2018) found lateral equilibrium positions of rigid particle train to be slightly closer to the channel wall than for a single particle. It is currently unclear how a train of soft particles under significant confinement would behave in relation to a single particle. Furthermore, we observed that the lateral equilibrium positions are independent of initial positions and the type of the particle interaction (data not shown).

For the stable pairs, where the axial distance $\delta x$ converges to a constant value after some time, Fig. 12(b) shows the axial equilibrium distance as function of $1/\text{La}$. We find that particles tend to form closer pairs when they are softer, and the lowest stable axial distance we observed is around $2.4a$, just above one particle diameter. Note that these particle pairs are also off-centre and on different sides of the channel centreline, so the actual distance between the particles is larger. We include the axial distance reported by Schaaf *et al.* (2019) for rigid pairs for comparison. The explanation of the gap between the rigid limit and the most rigid particle we have simulated is currently unclear. The axial distance could be strongly sensitive to mild particle deformation ($1/\text{La} < 0.01$), and future investigation of this parameter range is recommended.

The focusing time and distance, the time or distance until the last occurrence of a particle outside its equilibrium position within a given tolerance, is an important parameter in many inertial microfluidic applications. Fig. 12(c) shows the focusing time of the leading particle in the lateral direction and the axial distance between the particles in a pair. Three tolerances have been selected for each direction as representative low, medium and high values. The magnitude of these tolerances are arbitrary and intended to demonstrate the general trend of lateral and axial focusing times. Note that the absolute tolerances in the lateral and axial directions are equal to reflect tolerances in practical applications. We find that the focusing time decreases slightly in the lateral direction for softer particles in the range La = 50–125. However, for soft particles below La = 50, the focusing time tends to increase, surpassing the focusing time of the most rigid particles investigated when La = 20. This general trend is repeated for the focusing time in the axial direction, with the softest particles having the longest focusing time. Given that the softest particles have lateral equilibrium positions closest to the channel centreline, where the axial velocity is larger, this also corresponds to the softest particles having the largest focusing distance.

Fig. 12(c) shows that the lateral focusing occurs before the axial focusing. We will return to this observation in section 4.4 where we hypothesise that the particles within a pair must reach their lateral equilibrium position before being able to find their axial equilibrium distance.

Summarising the results so far, we have observed a number of 'rules' that all simulated particle pairs obey for the chosen value of Reynolds number (Re = 10) and particle-channel confinement ($a/h = 0.4$) and the range of Laplace numbers and initial positions investigated:

1. For all partially stable pairs, both particles are located at the channel centre.

2. For all stable pairs, both particles are off-centre and on different sides of the centreline and with essentially identical $\|z_{\text{eq}}\|$, *i.e.* $z_{\text{eq}}^{\text{lag}} \approx -z_{\text{eq}}^{\text{lead}}$.

3. A critical Laplace number exists for a given confinement ratio where particles with $\text{La} \leq \text{La}_{\text{cr}}$ always migrate to the channel centre and never form stable pairs.

These observations suggest that hydrodynamic particle interactions are fundamentally different for on-centre and off-centre pairs. Our findings, therefore, have important implications for microfluidic applications that rely on well-tuned axial distances between particles.

Finally, we observed that focusing of the lateral position occurs before stabilising the axial distance. We propose a high-level breakdown of the formation of stable particle pairs, shown schematically in Fig. 13. We distinguish between

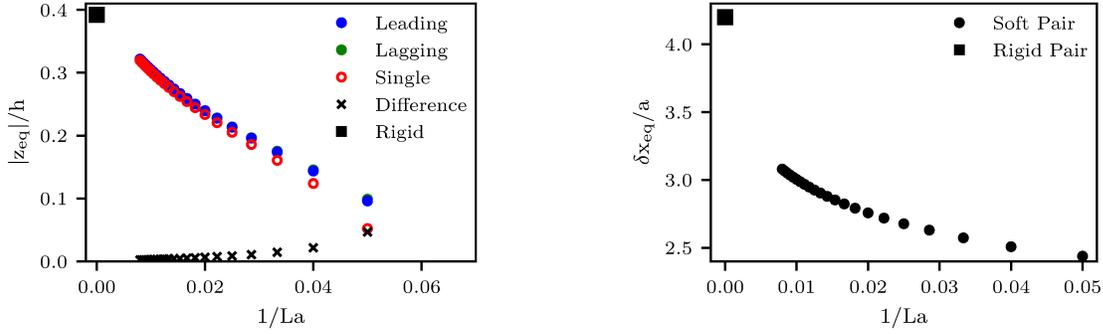

(a) Lateral equilibrium positions  (b) Axial equilibrium distances

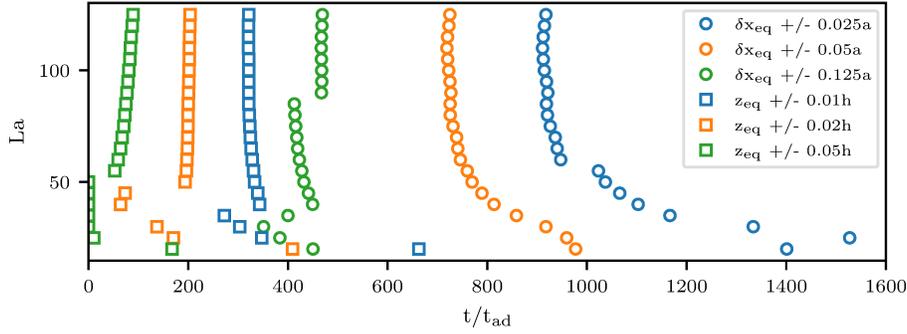

(c) Focusing time of particles with varying Laplace number for lateral equilibrium position and axial distance.

Figure 12: (a) Lateral equilibrium positions for leading particle (blue), lagging particle (green), and a single particle under the same conditions (red). Note that leading and lagging particle have essentially the same lateral equilibrium positions and green markers are not always visible. Crosses indicate the difference between the lateral positions of the leading particle in a pair and the single particle. (b) Axial equilibrium distance. (c) Focusing times for the leading particle in a pair in the lateral direction and for the axial distance. The focusing time is defined as the time until the last occurrence of the position/distance being outside its equilibrium value ± the specified tolerance. Three different tolerances are included to highlight the general trends.

zeroth-, first- and second-order effects. Zeroth-order effects are those that occur without particles present, *i.e.* the flow field development. First-order effects denote the behaviour of single particles that cannot be explained by the unperturbed flow field alone, in particular the single-particle lateral migration. Second-order effects are caused by the interaction of two particles, including the oscillation of the lateral position and the stabilisation of the axial distance between particles. Our results reveal that second-order effects strongly depend on the first-order effects while the presence of a second particle has only minor consequences for the lateral motion of the other particle. The notion of second-order effects depending on first-order effects have been observed experimentally; Gao *et al.* (2017) found that particle trains only begin to form once particle reach their lateral equilibrium positions. As a result, microfluidic designers must consider the key parameters of the application carefully. Devices that are used for cytometry rely on axial particle ordering (second-order) and will require a different quality of consideration than devices for particle separation (mostly relying on first-order effects).

### 4.4 Effect of Initial Position on Lateral and Axial Particle Migration

To better understand how stable pairs form, we analyse the oscillations of captured pairs when initial positions are varied. We first investigate the effect of initial axial distance $\delta x_0$. Fig. 14 shows the the lateral positions of the leading and lagging particle and axial distance in time for initial distances in the range $[3a, 11a]$. The Laplace number is set to La = 36 so that the lateral equilibrium position is equal to the initial position and the time development is mostly caused by axial rearrangement. Under these conditions, lateral oscillations are small (Fig. 14(a)); they arise from the flow field perturbations caused by the axially approaching particles (Fig. 14(b)).

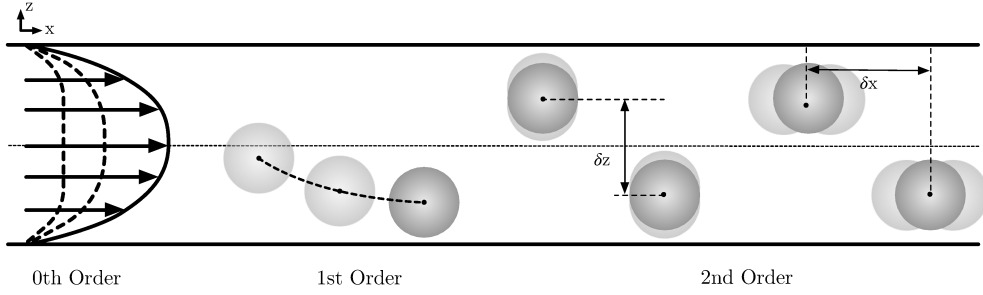

Figure 13: Contributions to the overall formation of stable particle pairs.

We identify the time $t_{\text{ax}}$ it takes for the axial distance to reach its first minimum as illustrated by vertical lines in Fig. 14(b). The resulting times $t_{\text{ax}}$ are shown in Fig. 15(a). It can be seen that the axial attraction time $t_{\text{ax}}$ has an exponential relationship with initial axial separation, with a behaviour $t_{\text{ax}}/t_{\text{ad}} \propto \exp(0.85 \delta x_0/a)$. Thus, even a small increase in initial particle separation might lead to pairs not forming before the particles have reached the device outlet.

Fig. 15(b) and (c) show the lateral position of the lagging particle and the axial distance shifted in time such that the first minima of $\delta x(t)$ coincide ($t \to t' = t - t_{\text{ax}}$). We find that the amplitudes of both the lateral and axial oscillations increase with $\delta x_0$, converging to a constant magnitude when $\delta x_0 \approx 7a$. These observations suggest two phases exist in the axial migration of particle pairs. The first phase at $\delta x \geq 7a$ is slow, while the second phase at $\delta x < 7a$ is faster and includes stronger particle interactions leading to oscillations in both lateral position and axial distance. Importantly, the data shows that the second phase is not strongly affected by the first phase. Only for cases where particles are initially closer than $7a$, the system is initialised directly in the second phase, and the outcomes strongly depend on the initial distance $\delta x_0$.

Next we investigate the effect of initial lateral position $z_0$. The parameters are La = 90 and $\chi = 0.4$ for which the lateral equilibrium position is $z_{\text{eq}} \approx \pm 0.3h$ and the axial equilibrium distance is $\delta x_{\text{eq}} \approx 3a$ for all cases investigated. Particles are released at $\delta x_0 = 3a$ and on different sides of the centreline, i.e. $z_0^{\text{lag}} = -z_0^{\text{lead}}$. Fig. 16(a) and (b) show the time evolution of the lateral position of both particles and their axial distance for a range of initial lateral positions, $[0.1h, 0.45h]$. Oscillations are visible in both the lateral position and the axial distance. The oscillation amplitude tends to be larger when the initial position $z_0$ is farther away from $z_{\text{eq}}$. As expected, for the special case $z_0 \approx z_{\text{eq}}$, there are virtually no oscillations since particles are initialised close to their equilibrium configuration. For the majority of cases, the initial lateral motion of both particles is towards the equilibrium position. However, lagging particles initially located between their equilibrium lateral position and the channel centreline behave differently: these particles first move away from their equilibrium position (towards the centreline) before changing direction and eventually converging at the equilibrium. Given that single particles in our simulations always move towards their equilibrium lateral position, this observation can be attributed to hydrodynamic particle-particle interactions.

The direction of the initial oscillation of the axial particle distance (Fig. 16(b)) differs either side of the lateral equilibrium position with growing amplitude as the difference between initial and equilibrium lateral position increases: when particles in a pair are initially closer to the centreline than their lateral equilibrium positions, particles first decrease their axial distance. Particles initially farther away from the centreline first increase their axial distance. The inset in Fig. 16(b) shows the two extreme cases with initial lateral positions of $0.1h$ and $0.45h$. Despite the amplitude of the first oscillation of both pairs being similar, the oscillation of the pair farthest away from the channel centreline damps more quickly than the pair closest to the centreline. We will explore the damping of the oscillations in more detail in Section 4.5.

We now consider the centre-of-mass behaviour of the particle pair. Fig. 16(c) shows the lateral centre of mass position versus the axial particle distance. Since particles are released with the same axial distance and at the same distance either side of the channel centre, the centre of mass always starts at the same point on the centreline (circle in Fig. 16(c)). The inset shows a zoomed area close to the initial and equilibrium positions. At early times, the centre of mass leaves the centreline, begins to oscillate and eventually spirals towards the equilibrium back on the centreline. The initial direction of the motion of the centre of mass depends on the initial lateral position of the particles. However, as the pairs approach the equilibrium positions at later times, all trajectories converge to a similar counterclockwise path. This convergence implies that the time history of the pair is eventually forgotten, which is in line with observations in Fig. 15 suggesting that the behaviour during the second phase is largely independent of that during the first phase of migration.

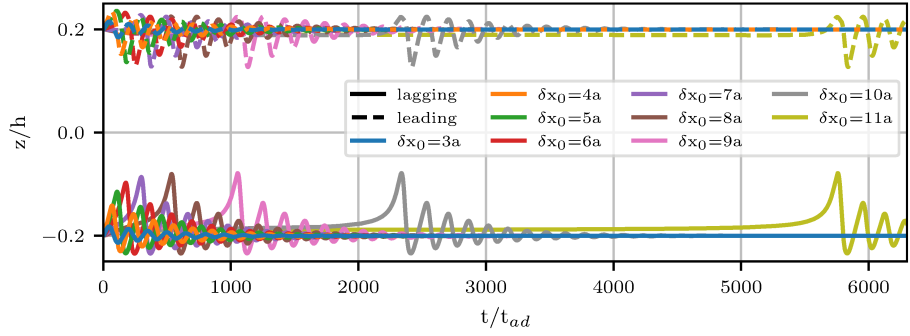

(a)

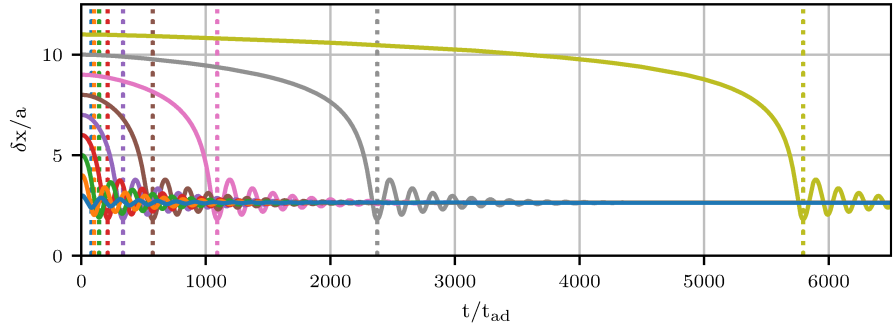

(b)

Figure 14: Time evolution of (a) lateral particle positions and (b) axial distance between particles at La = 36 and $\chi = 0.4$. Initial lateral positions are set to $z_{\text{eq}}$. The initial axial distance is increased until a captured pair does not form before $t/t_{\text{ad}} \geq 10,000$.

Finally, Fig. 16(d) shows the time evolution of the half-lateral distance between particles in a pair, compared to the lateral position of a single particle with the same initial position. Generally, both types of curves follow the same trend, *i.e.* to first order, particles in a pair and single particles behave similarly. Interestingly, particles in a pair reach their lateral equilibrium distance faster than the single particle converges to its lateral equilibrium position when particles are initially close to the centreline. This difference could be caused by the initial proximity of both particles which repel each other hydrodynamically, hence accelerating the initial lateral migration away from the centreline. Particles in a pair that are initially farther away from each other experience a weaker hydrodynamic repulsion and behave nearly like single particles. Another important observation is that the lateral oscillations seen in Fig. 16(a) are nearly completely absent in Fig. 16(d). On long time scales — of the order of hundreds of advection times — particles can change their lateral distance and migrate towards their equilibrium; but on the shorter time scale of the lateral oscillations — tens of advection times — particles do not oscillate relatively to each other. We hypothesise that a fast oscillation of the lateral particle distance is suppressed by the liquid between the particles: a fast lateral oscillation would require liquid being repeatedly squeezed in and out of the gap between the particles, which is energetically unfavourable. Therefore, the particles' lateral positions are hydrodynamically coupled on a short time scale, which also explains why in some cases one particle in a pair initially moves away from its equilibrium position (Fig. 16(a)).

Taking all observations from Fig. 16 together, we can now explain the counterclockwise sense of the oscillation seen in Fig. 16(c). During periods when the lagging particle is closer to the centreline than the leading particle, the lateral

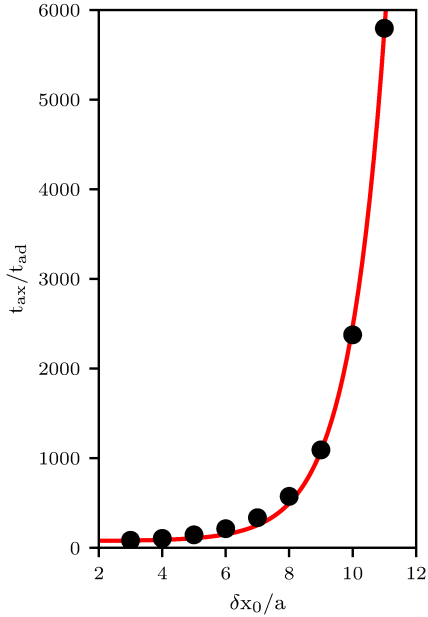

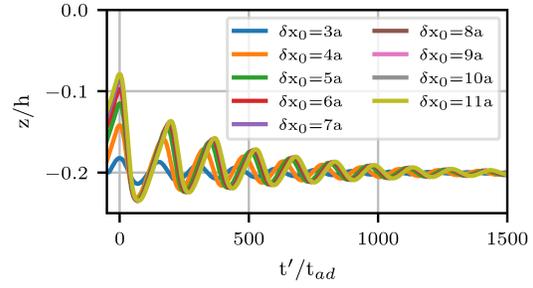

(b) Lateral position of the lagging particle versus shifted time $t' = t - t_{\text{ax}}$.

(a) Axial attraction time as function of initial axial distance.

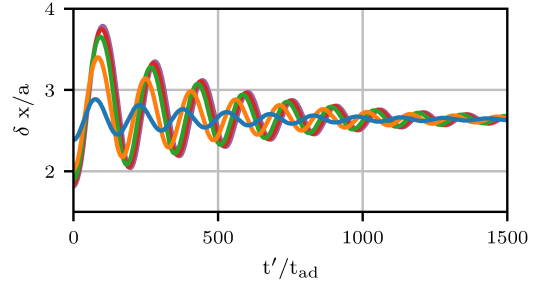

(c) Axial distance versus shifted time $t' = t - t_{\text{ax}}$.

Figure 15: (a) Axial attraction time $t_{\text{ax}}$ at which the axial distance reaches its first minimum as denoted by vertical lines in Fig. 14(b). The solid line is an exponential $\propto \exp(0.85 \delta x_0/a)$. (b) Time evolution of lateral position of the lagging particle for different initial axial distances. (c) Time evolution of axial distance for different initial axial distances.

centre of mass position is positive and the lagging particle is able to catch up with the leading particle ($\delta x$ decreases) since the lagging particle is exposed to a faster portion of the flow field. Once particles approach axially, the gap between both particles is decreasing, and liquid needs to be pushed out of the gap, causing a repulsion of the particles. This repulsion force affects both particles differently. The particle currently closer to the centreline experiences a smaller shear gradient force and can move more easily away from the centreline than the particle currently in a higher shear-gradient region. As a consequence and supported by particle inertia, the lateral centre of mass location crosses the centreline and turns negative, and the leading particle becomes faster, therefore increasing the axial distance $\delta x$. Upon increasing the axial distance, the gap grows and liquid needs to move into the gap, therefore causing particle attraction. The process then continues with swapped roles of both particles and repeats itself. While inertia drives this oscillation, viscosity causes the damping and eventual convergence to the equilibrium state.

Having identified the importance of oscillations in the formation of a pair, we now focus our investigation on the dynamics of these oscillations and their effect on focusing time.

### 4.5 Capture Oscillation Dynamics and Focusing Time

Previous studies have identified the existence of damped oscillation of the relative distance between captured particles (Schaaf *et al.*, 2019; Hu *et al.*, 2020; Udono, 2020). To understand better how stable pairs form, we analyse the oscillations of the axial distance between particles when different parameters are varied: Laplace number, initial lateral position $z_0$ and initial axial distance $\delta x_0$. Note that the trajectories analysed in this section have already been presented in Section 4.3 for variable Laplace number and Section 4.4 for variable initial positions. We evaluated the damped frequencies by measuring the time periods between oscillation peaks, similar to the earlier analysis of rigid particle pairs (Schaaf *et al.*, 2019). Fig. 17(a) shows that the oscillation frequency increases with La until about La > 60 beyond which the frequency becomes approximately constant. The oscillation frequency of the softest particle pair studied is about 50% smaller than that of the most rigid particle pair studied. The inset in Fig. 17(a) reveals the results of a fast

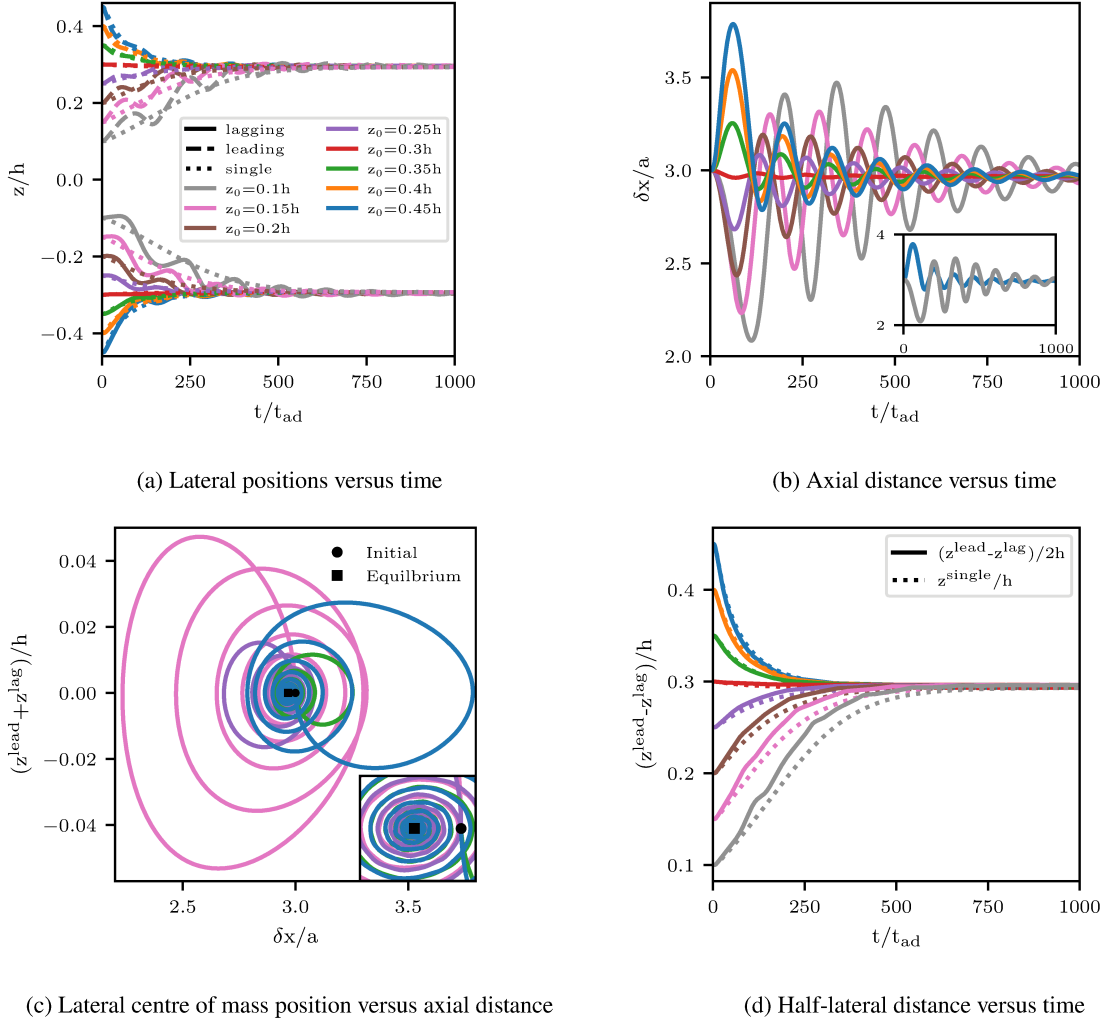

(a) Lateral positions versus time

(b) Axial distance versus time

(c) Lateral centre of mass position versus axial distance

(d) Half-lateral distance versus time

Figure 16: Time evolution of (a) lateral particle position and (b) axial distance at La = 90, $\chi = 0.4$ and $\delta x_0 = 3a \approx \delta x_{\mathrm{eq}}$. Initial lateral positions obey $z_0^{\mathrm{lag}} = -z_0^{\mathrm{lead}}$ and are varied in the range $[0.1h, 0.45h]$. (c) Lateral centre of mass position of the pair versus axial distance; the inset shows the zoomed area close to initial and equilibrium positions. (d) Time evolution of half-lateral distance between particles compared to trajectories of single particles. Note that a reduced selection of initial positions are included in (c) to improve readability. The line colours in all panels correspond to the legend in (a).

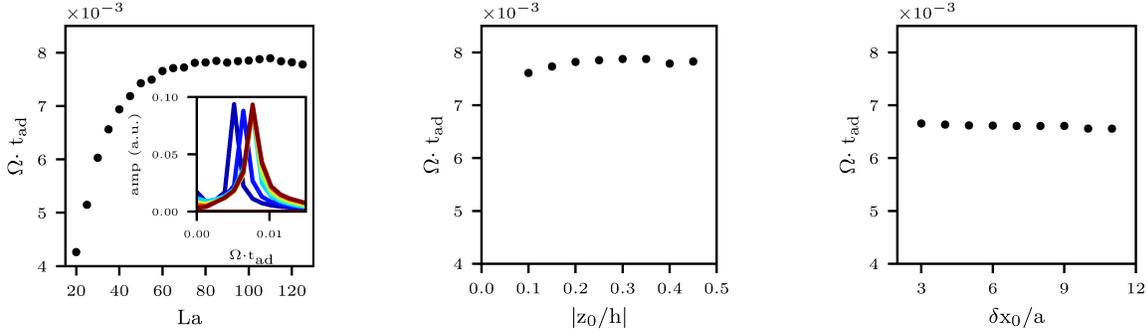

(a) Frequency as function of particle softness

(b) Frequency as function of initial lateral position (La = 90)

(c) Frequency as function of initial axial distance (La = 36)

Figure 17: Oscillation frequency $\Omega$ as function of (a) Laplace number, (b) initial lateral position (La = 90) and (c) initial axial distance (La = 36). The inset in (a) shows results from a fast Fourier transform of the time data for a selection of Laplace numbers (colours correspond to those in Fig. 10(a)).

Fourier transform of the time data, confirming the findings in the main graph.

We investigated the influence of initial configuration on the resulting oscillation frequency during pair capture. Fig. 17(b) and (c), respectively, show the obtained frequencies as function of initial lateral position for La = 90 and initial axial distance for La = 36. Fig. 17(b) corresponds to the cases in Fig. 16, and Fig. 17(c) corresponds to the cases in Fig. 14. We see a slight increase in frequency with lateral position and a slight decrease in frequency with initial axial distance. However, the variation in both relationships is over an order of magnitude smaller than that of the variation with Laplace number. We conclude that particle softness is the key determining factor for the oscillation behaviour, and initial positions have a negligible effect on the late-stage capturing dynamics where oscillations are strong. These findings support our hypothesis that the particle dynamics during the second phase of migration are largely independent of the initial details and the first phase of migration.

Next we analyse the damping of the oscillations of the axial distance. To obtain the damping coefficient $\gamma$ for each case, we fitted decaying exponentials of the form $A \exp(-\gamma(t - t_0)) + \delta x_{eq}$ to the maxima and minima of the time data $\delta x(t)$. Since the initial axial distance $\delta x_0$ generally has a mismatch with the axial equilibrium distance $\delta x_{eq}$, the oscillations of $\delta x(t)$ show pronounced transients at early times. To improve results, we ignored the first two periods of each curve in the fitting process. The inset of Fig. 18(a) illustrates our procedure. Fig. 18 shows the results for the damping ratio $\gamma$; panels (a), (b) and (c) correspond to those in Fig. 17. Fig. 18(a) reveals that the damping ratio increases with La. Unlike the frequency $\Omega$ in Fig. 17(a), $\gamma$ keeps increasing beyond La = 60 and does not seem to converge to a 'rigid limit' in the range of Laplace numbers investigated. However, Fig. 18(b) and (c) show that the damping ratio — as the frequency in Fig. 17(b) and (c) — is essentially independent of initial lateral position and initial axial distance. The behaviour of the damping coefficient provides further support to the importance of particle softness to the oscillation dynamics, while details of the initial conditions are less relevant.

A key parameter in the formation of particle pairs is the focusing time for both the lateral position and the axial distance. In Section 4.3, we presented the dependence of both focusing times on La. We showed that particles are focused laterally before the axial distance equilibrates. Fig. 19 reveals the dependence of focusing time on particle softness, initial particle position and initial axial distance with moderate tolerance ($\delta x_{eq} \pm 0.05a$ and $z_{eq} \pm 0.02h$). For clarity, Fig. 19(a) contains the relevant subset of data from Fig. 12(c). As expected, Fig. 19(b) and (c) show that the initial conditions have a strong effect on the focusing time. While focusing times are minimal when particles are initialised close to their equilibrium lateral position ($z_{eq} \approx 0.3h$ in this case), focusing times increase with $|z_0 - z_{eq}|$ (Fig. 19(b)). Interestingly, focusing times become particularly large when particles are initially close to the centreline; probably because shear gradients are smaller in this region. This trend follows the change in oscillation amplitudes observed in Fig. 16. Likewise, focusing times are smallest when the initial axial distance matches the equilibrium distance ($\delta x_{eq} \approx 3a$ in this case), while the focusing time increases strongly with $\delta x_0$, in particular for $\delta x_0 > 9a$ (Fig. 19(c)). At lower values of $\delta x_0$, the increase in focusing time is much more gradual, perhaps offsetting the decrease in axial distance with the changing oscillation dynamics that are constant at larger initial axial distance (Fig. 15(c)).

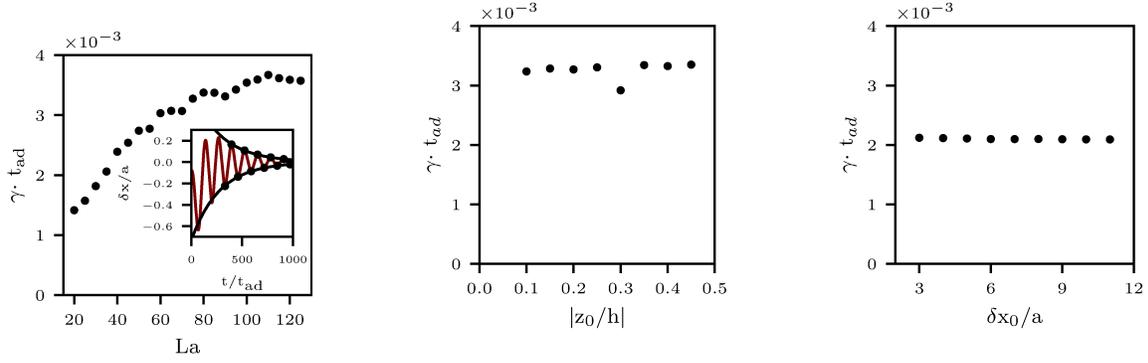

(a) Damping as function of particle softness  
(b) Damping as function of initial lateral position (La = 90)  
(c) Damping a function of initial axial distance (La = 36)

Figure 18: Damping ratio $\gamma$ as function of (a) Laplace number, (b) initial lateral position (La = 90) and (c) initial axial distance (La = 36). The inset in (a) depicts the process of obtaining the damping coefficient $\gamma$ (see main text for details).

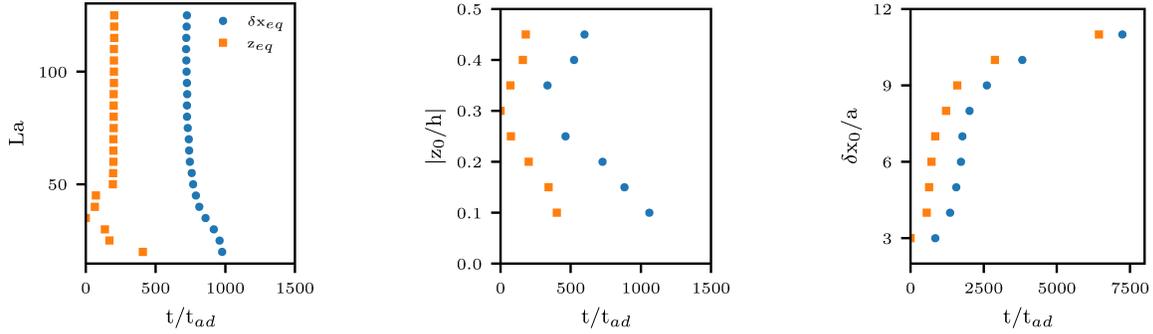

(a) Focusing time as function of particle softness  
(b) Focusing time as function of initial lateral position  
(c) Focusing time as function of initial axial distance

Figure 19: Focusing time for lateral position of the leading particle and the axial distance as function of (a) Laplace number, (b) initial lateral position (La = 90) and (c) initial axial distance (La = 36).

Our analysis shows that both the initial positions and particle softness play an important role in the dynamics of stable pair formation. Particle softness dominates the characteristics of trajectory oscillations while initial positions play a larger role in determining focusing time. We conclude that designers of inertial microfluidic devices must also consider the particle positions prior to interaction, with device performance optimised through consideration of particle softness.

## 5 Conclusions

The formation of stable particle pairs is the foundation for many applications within the field of inertial particle microfluidics, including particle focusing, cytometry and separation. Identifying the conditions under which stable pairs form is crucial to improving the design process and performance of some inertial microfluidic devices. Despite most particles processed in microfluidic devices being soft biological cells, the majority of numerical investigations to date have concentrated on the modelling of rigid particles. It is known that particle deformability affects the migration behaviour under the influence of inertia. Here, we have investigated the effect of particle softness on the formation and stability of particle pairs in straight channels under moderate inertial flow conditions.

We used an in-house lattice-Boltzmann-immersed-boundary-finite-element solver to simulate single and pairs of soft capsules in channel flow under the influence of fluid inertia at channel Reynolds number 10 and particle-to-channel confinement of 0.4. The code has been benchmarked against previously published numerical results involving a pair of capsules interacting in inertial shear flow and a single soft particle migrating in a channel.

There are several important results:

1. We first investigated the general effect of particle softness, characterised by the Laplace number, and initial particle position for a pair of equally soft particles. We observed two new trajectory types termed *Swap & Capture* and *Pass & Capture* that appear in some cases and have not been found for rigid particle pairs in the same geometry in earlier studies. Particle softness was found to increase the likelihood of a captured particle pair forming.

2. We observed that particle pairs that migrate to the channel centre are stable only in the lateral direction, but not in axial direction (partially stable pairs). In contrast, particle pairs that migrate to an off-centre lateral equilibrium position form pairs that are stable in both the lateral and axial directions. We found that particle softness seems to affect the pair stability only through the resulting lateral equilibrium position. Generally, the lateral equilibrium positions of soft particles in a pair are nearly the same as those of single particles under the same conditions. Importantly, our simulations show that the stabilisation of the axial particle distance occurs only after particles have migrated laterally. Our findings suggest that the observations can be classified as zeroth-, first- and second-order effects which refer to effects caused by the background flow field, the behaviour of a single particle, and the interaction of two particles, respectively.

3. The formation of a stable pair consists of two phases: an early and a late phase. During the early phase, particles migrate laterally and approach each other until the axial distance becomes small. The later phase involves a spiralling motion leading to a converged state. We found the late phase to be largely independent of the first phase, as long as the initial conditions lead to the eventual formation of a stable pair. During the spiralling motion towards the converged state, particles are tightly coupled through hydrodynamic interactions. The spiralling motion is driven by an interplay of flow-induced lift, particle inertia and viscous dissipation. Importantly, the pair formation time grows exponentially with the initial axial distance of the particles.

4. The particle oscillations during the late phase of pair formation are characterised by their frequency and damping coefficient. Both quantities increase with Laplace number until a rigid particle limit is reached, but they are largely independent of initial conditions. Finally, we show that the pair formation time is determined by the initial conditions while particle softness has only a mild influence.

Through our investigation, we have identified new physical effects in the formation of stable particle pairs that are softness-dependent. These effects should be considered in the design of inertial microfluidic devices, given that soft particles are used extensively in real-world applications. Since axial stabilisation comes after lateral migration, our observations put lower limits on the required time and distance necessary to generate stable pairs. Furthermore, devices relying on appropriate axial spacing require more consideration than devices that focus or separate particle by lateral position. Our findings could have significant ramifications for applications where a constant and reliable axial distance between particles is required, such as flow cytometry.

Future work could include a detailed analysis of the role of the flow field in the formation of pairs and trains of particles and the development of reduced-order models to predict pair formation without the need for resolved simulations. Since many inertial microfluidic applications involve heterogeneous particle mixtures, an investigation of heterogeneous pairs and trains over a larger range of Laplace numbers would provide further insight.

## Acknowledgements

We thank Christian Schaaf and Holger Stark for providing numerical benchmark data and, along with Prosenjit Bagchi and Erich Essmann, their fruitful discussion. This work used the Cirrus UK National Tier-2 HPC Service at EPCC (http://www.cirrus.ac.uk). TK received funding from the European Research Council (ERC) under the European Union's Horizon 2020 research and innovation program (803553).